\documentclass{article}

\usepackage[left=1in, right=1in, top=1in, bottom=1in]{geometry}
\usepackage{amsmath}
\usepackage{amssymb}
\usepackage{graphicx}
\usepackage{color}
\usepackage{aas_macros}
\usepackage[authoryear,semicolon]{natbib}
\usepackage[citecolor=blue,linkcolor=blue,colorlinks]{hyperref}
\usepackage{hyperref}
\usepackage{tikz}
\usetikzlibrary{shapes,arrows,positioning,decorations.pathreplacing}
\usepackage{authblk}
\usepackage{textpos}

\author[1]{\Large Kanhaiya L. Pandey\thanks{kanhaiya.pandey@iiap.res.in}}
\author[1]{\Large A. Mangalam\thanks{mangalam@iiap.res.in}}
\affil[1]{\large Indian Institute of Astrophysics, Bengaluru, India}

\title{Role of primordial black holes in the direct collapse scenario of supermassive black hole formation at high redshifts}

\date{}

\begin{document}

\maketitle

\begin{textblock*}{120mm}(.4\textwidth,-7cm)
{Accepted for publication in Journal of Astrophysics and Astronomy}
\end{textblock*}

\begin{abstract}

In this paper, we explore the possibility of accreting primordial black holes as the source of heating for the collapsing gas in the context of the direct collapse black hole scenario for the formation of super-massive black holes (SMBHs) at high redshifts, $z\sim 6-7$. One of the essential requirements for the direct collapse model to work is to maintain the temperature of the in falling gas at $\approx 10^4$ K. We show that even under the existing abundance limits, the primordial black holes of masses $\gtrsim 10^{-2} \ {\rm M}_\odot$, can heat the collapsing gas to an extent that the ${\rm H}_2$ formation is inhibited. The collapsing gas can maintain its temperature at $10^4$ K till the gas reaches a critical density $n_{crit} \approx 10^3$ cm$^{-3}$, at which the roto-vibrational states of ${\rm H}_2$ approaches local thermodynamic equilibrium and ${\rm H}_2$ cooling becomes inefficient. In the absence of ${\rm H}_2$ cooling the temperature of the collapsing gas stays at $\approx 10^4$ K even as it collapses further. We discuss scenarios of subsequent angular momentum removal and the route to find collapse through either a supermassive star or a supermassive disk.

\end{abstract}

{\bf keywords- }{cosmology: theory; cosmology: dark ages, reionization, first stars; quasars: supermassive black holes.}

\section{Introduction}
A number of bright quasars have been found at $z \gtrsim 7$ \citep{2011Natur.474..616M,2015Natur.518..512W}. 
These high energy sources are believed to be powered by gas accretion onto a massive ($\gtrsim 10^9 {\rm M}_\odot$) compact region, suggesting a presence of super-massive black hole at the centre of each of these quasars. The existence of such SMBHs with masses $\approx 10^9 \ {\rm M}_\odot$ at the time when the Universe had just spent $\approx$ 5-6\% ($\approx 700$ Myr) of its total age, raises a big question of how such massive black holes could have been formed so early in the Universe. 
There are three main proposals in the literature for black hole genesis, all of which have caveats.
The first one starts with star clusters, which if compact enough, can suffer
runaway collisions that could lead to a single collapsed star with a mass of about $100 M_\odot$. Simulations of this scenario suggest that black hole seed formation is not sufficiently efficient; besides this, it is not clear how these dense star clusters are formed in the first place \citep{2012RPPh...75l4901V}.

Another idea is that seed black holes are formed from single massive stars; in case of Pop I and II stars, the presence of metals enhances opacity and limits the mass of the star that collapses to black holes due to excessive stellar winds and mass loss. So, it is the first generation of stars (the Pop III stars) that would leave behind black holes of masses of at least $\approx 100 \ {\rm M}_\odot$, early enough by the  redshifts as high as $z \sim 20-25$ \citep{2002Sci...295...93A,2002ApJ...564...23B,2007ApJ...654...66O}. However, growing these seed black holes by gas accretion at Eddington rate would take $\approx 10^8$ years to reach a mass of around $10^9 \ {\rm M}_\odot$ \citep{1964ApJ...140..796S}, which is much longer than the time available to reach it by $z \sim 7$. Although, one can assume a sufficiently efficient gas accretion to make it work, but it turns out that the accretion has to be highly super-Eddington \citep{2015MNRAS.451.1964S}. Such efficient growth by accretion is very unlikely due to many feedback effects such as radiation pressure and photoionization heating, which can lead to a severe disruption of the accretion flow. 
Also, neighboring Pop III stars in the vicinity of the seed black hole, and the seed black hole itself may ionize the surrounding gas causing suppression of subsequent accretion onto the progenitor seed black hole \citet{2012MNRAS.426.1159S}. Merging of the black holes can also play an essential role in the formation of the SMBHs, but it turns out to be difficult as well, typically the number of mergers required is too high. Very frequent halo mergers would result in the formation of small N-body systems of BH seeds, which would be prone to SMBH slingshot ejections from the DM halo \citep{2004ApJ...613...36H, 2007ApJ...663L...5V}.

An appealing alternative class of models called the direct collapse black hole (DCBH) model for the formation of SMBHs involves formation of a SMBH seed of mass $\sim 10^5 \ {\rm M}_\odot$, by the redshift of $z \sim 10-20$ through a rapid collapse of primordial (metal-free) gas in a high density, high inflow rate environment in which star formation is suppressed \citep{2002ApJ...569..558O, 2003ApJ...596...34B, 2005ApJ...633..624V, 2006MNRAS.370..289B}. This model has gained an increasing interest following a recent observation of highly luminous Ly$\alpha$ emitter galaxy (CR7) at $z \sim 6.6$. The observational features of the galaxy CR7 (such as absence of metal lines, presence of Ly$\alpha$ and He${\rm \scriptscriptstyle II}$ 1640 \r{A} line with a large +160 km/s offset) put together fits very well with the DCBH modeling of the data, indicating that the object CR7 contains a direct collapse black hole in making at its centre \citep{2016MNRAS.460.3143S}.

In the context of DCBH formation model, the issues anticipated in the case of Pop III seed scenario do not play any important role. In order to make this model work, the collapsing gas must avoid fragmentation, shed angular momentum efficiently, and collapse rapidly. For these conditions to be met, the gas needs to keep its temperature high enough ($\sim 10^4$ K). This kind of rapid gas collapse can be expected in relatively massive dark matter halos with virial temperature $T_{vir} \gtrsim 10^4$ K. However, recent numerical simulations \citep{2010MNRAS.402.1249S} find that the collapsing gas in such halos forms ${\rm H}_2$ efficiently and cools to temperatures $\lesssim \ 300$ K. 

The ${\rm H}_2$ cooling is an important concern in the DCBH scenario. However, if the in falling gas is exposed to an intense Lyman-Werner (band near photon energy $\sim 12$ eV) UV flux coming from sources in its vicinity, the formation of ${\rm H}_2$ can be suppressed (either by directly photo-dissociating ${\rm H}_2$ or by photo-dissociating intermediary ${\rm H}^-$) and thus the ${\rm H}_2$ cooling can be avoided. The critical LW flux needed for this to happen is much higher, $J^{LW}_{crit} \gtrsim 10^{2-5}$, \citep{2010MNRAS.402.1249S,2001ApJ...546..635O,2003ApJ...596...34B} than the expected level of cosmic UV background at those high redshifts, eg. $J^{LW} \sim 1$ at $z \sim 10$ \citep{2009ApJ...694..879T} \footnote{$J^{LW}$ is in units of $10^{-21}$ erg s$^{-1}$ Hz$^{-1}$ str$^{-1}$ cm$^{-2}$.}. However, the flux seen by an individual halo may fluctuate by many orders of magnitude depending on the distance of the source from the halo, a possible high enough LW flux coming from nearby source can potentially contaminate the collapsing halo with metals and thus making the collapsing gas cool much faster and consequently fragment. 

A lot of scenarios have been studied to circumvent the H$_2$ cooling in the context of SMBH formation models, such as \cite{2010ApJ...721..615S} and the references therein. We explore the possibility of accreting primordial black holes as the heating source. The idea that primordial black holes (PBHs) might comprise some or all of the dark matter has been around for many decades. 
Also this hypothesis has been tested by various observations such as, extragalactic $\gamma$-ray background, OGLE (The Optical Gravitational Lensing Experiments)-I,II (MACHO and EROS), III \& IV, microlensing from Kepler,  Eridanus-II star cluster, and CMB experiments such as WMAP and Planck (\citet {2016PhRvD..94h3504C} and references therein, \citet{2008ApJ...680..829R,2017PhRvD..95h3006C}), which gave rise to strong bounds on PBH abundances in various mass windows. Even with the latest experiments there are certain mass ranges for which $f_{\rm PBH}$ ($= \Omega_{\rm PBH}/\Omega_m$) could be as high as 10\%. 

\begin{figure}[!ht]
\begin{center}
\includegraphics[width=1.0\columnwidth]{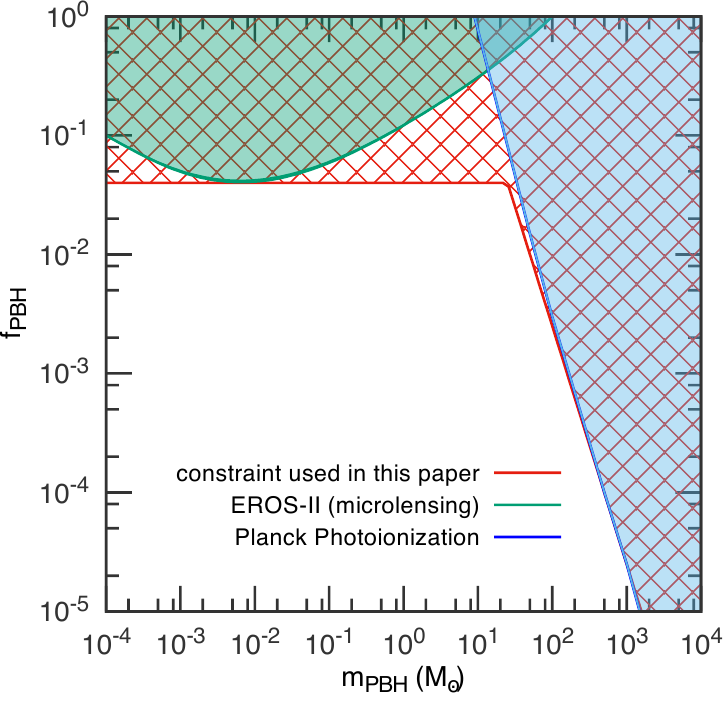}
\caption {Constraints on $f_{PBH} = \Omega_{PBH}/\Omega_m$ coming from micro-lensing from EROS-II \citep[monochromatic constraints]{2017PhRvD..96b3514C}, Planck \citep[assuming photoionization through PBH accretion]{2017PhRvD..95d3534A} and the constraint which we have used in this paper, are shown above.}
\label{fig:const}
\end{center}
\end{figure}

We use the following form of the $\Omega_{\rm PBH}$ for our calculation. This particular form is a simpler adaptation of the constraints on the PBH abundances in the mass range of the interest ($10^{-4} {\rm M_\odot} {-} 10^{4} {\rm M_\odot}$), see Figure~\ref{fig:const}.
\begin{equation}
\Omega_{\text{PBH}} = 
\begin{cases} 
25 \times \left(\frac{\rm m_{PBH}}{\rm M_\odot}\right)^{-2} \Omega _m, & {\rm m_{PBH}}>25 \ {\rm M_\odot} \\
0.04 \ \Omega _m, & {\rm m_{PBH}}<25 \ {\rm M_\odot}
\end{cases}
\label{eq:1}
\end{equation}
where ${\rm m_{PBH}}$ is the mass of the individual PBH, $\Omega_m$ is the cosmological dark matter density parameter.

\begin{sloppypar}
We assume that a portion ($f_{\rm PBH}$) of dark-matter is made up of primordial black holes of a certain mass, which are uniformly distributed inside the dark matter halo. We discuss some findings on the possible cosmological evolution of primordial black holes \citep{2017JHEAp..13...22R} using a detailed Eddington limited cosmic accretion history of all components of baryons, dark matter (WIMPs) and radiation:
\begin{itemize}
\item
PBHs with a wide range of masses could have been formed in the very early Universe as a result of the direct collapse of large over-densities \citep{1971MNRAS.152...75H,2017PhRvD..95h3006C}.
\item
A PBH with mass $<10^{14}$ g will not accrete radiation or matter at a rapid enough rate and will evaporate due to Hawking radiation.
\item
A PBH with initial mass (in g) in the range $10^{15} < M_{PBH} < 10^{28}$ neither evaporates or accretes significantly over Hubble time. 
\item
A PBH with initial mass (in g) in the range $10^{28} < M_{PBH} < 10^{29}$ accretes
significantly during the era after recombination and before thermalization of baryons with radiation, $z_{th}\sim 174 < z < z_{rec} \simeq1100$ at Eddington limited  Bondi rates. The mass changes by an order of magnitude over this time
and subsequently increases only by a small fraction.
\item
It is unlikely for PBHs to grow to large black holes by $z \sim 6$ by Eddington
limited accretion.
\end{itemize}
With this in mind, we now turn our attention to the extent to which
the radiation from these accreting PBH can heat the ambient Universe in the
from post-recombination era to $z \sim 7$.

The $f_{\rm PBH}$ is a function of the mass of the PBH (${\rm m_{PBH}}$) and is taken from the most recent constraints on the PBH abundance. We use a monochromatic abundance limit on PBH of masses around 1 ${\rm M}_\odot$, as suggested by \citet{2017PhRvD..95d3534A}. These limits are for accreting primordial black holes using the recent Planck CMB temperature and polarization data. They have considered the effect of the radiation coming from the PBH accretion on the collisional ionization and the photoionization in the ambient medium. The constraints coming from the photoionization considerations are stronger, and to be conservative we have taken the stronger constraint on $f_{PBH}$ quoted by them. For masses below 1 M$_\odot$ we use the EROS-II microlensing constraints \citep{2007A&A...469..387T}. 
We find that even under the given constraints on the PBH abundances [though we use a little stronger constraints to be more conservative, Equation~(\ref{eq:1}), Figure~\ref{fig:const}], accreting PBHs of masses $\gtrsim 10^{-2} \ {\rm M}_\odot$, uniformly distributed in a dark matter halo, can provide enough heating for the collapsing gas to suppress the ${\rm H_2}$ formation and hence the gas can avoid the ${\rm H_2}$ cooling.

We would like to emphasize here that, unlike the case of the suppression of H$_2$ abundance through LW flux, where the H$_2$ molecules are directly gets destroyed by the radiation, in this scenario, the suppression of H$_2$ abundance is due to the suppression in the H$_2$ formation rate because of higher temperature of the gas. Also in this scenario, there is no issue of contamination of the gas with heavy metal coming from nearby LW sources (stars).
\end{sloppypar}

\section{Thermodynamical evolution in DCBH scenario with accreting PBH heating}
\begin{sloppypar}
We carry out a study of a collapsing gas into the potential well of dark matter halo of mass ${\rm M}_h = 10^8 \ {\rm M}_\odot$. The initial condition is set to be as $\delta_i = 0.04$ at $z=1000$, (a 6-7 $\sigma$ fluctuation with the collapse redshift $z_{col} \sim 25$ ). Initially, the gas traces the evolution of the dark matter till the dark matter gets virialized and turns into a dynamically stable halo. After dark matter virialization, gas collapses into the gravitational potential of the dark matter halo as it loses its pressure support through various cooling processes.
For the thermal evolution of the collapsing gas, we use the prescription given in the \citet{2017PhRvD..95h3006C}, with an additional term in heating given by accreting primordial black holes ($dE_{\rm PBH}$/dVdt), and we also include atomic H and molecular H$_2$ cooling  (Equation~\ref{eqn:dedvdt}). For H and H$_2$ cooling we follow the prescription given in \citet{2010ApJ...721..615S}.

\subsection{\underline{Dynamical evolution of the collapsing gas}}

We use the simple spherical top-hat model for the collapse of the matter till the dark matter halo reaches virialization. We assume that till this stage the dark matter and the baryons follow each other. The evolution of the radius of the encompassing shell in comoving coordinates is computed using the following equations: 
\begin{flalign}
r(z) &= r_i \frac{1+\delta_i}{2\delta_i} (1-\cos\eta(z)) \\
\frac{t(z)}{t_i} &= \frac{\eta(z) - \sin\eta(z)}{\eta_i - \sin\eta_i}
\end{flalign}
where $\cos\eta_i \equiv 1-2\delta_i/(1+\delta_i)$ and 
\begin{flalign}
r_i &= \left(\frac{3 M_i} {4\pi (1+\delta_i) \rho_i^{crit}} \right)^{1/3} \\
\rho_i^{crit} &= \frac{3 {H_0}E(z_i)}{8\pi{\rm G}} \\
t(z) &= \frac{1}{H_0}\int_z^\infty \frac{dz}{(1+z) E(z)}
\end{flalign}
where $E(z) = (\Omega_m (1+z)^3 + \Omega_\Lambda)^{1/2}$ and $M_i = M_h(\Omega_m/\Omega_c)$. 
$\Omega_m, \ \Omega_c$ and $\Omega_b$ are respectively density parameter for the total matter, dark matter and baryons ($\Omega_m=\Omega_c+\Omega_b$). Since at high redshifts, the Universe was matter dominated, we can safely use the above mentioned spherical top hat equations for the dynamical evolution of the dark matter halo. The gas is assumed to follow the dark matter halo till virialization.
After virialization, the baryonic gas falls into the virialized dark matter halo by losing its thermal energy due to inverse-Compton, atomic H, and molecular H$_2$ cooling. We evolve gas density after dark matter virialization until it reaches a number density of $\approx$ a few 1000 per cc, using simple a gravitational collapse given by
\begin{flalign}
\frac{d^2 r}{dt^2} &= -\frac{G M(r)}{r^2} \\
M(r) &= M_h \left(\frac{\Omega_b}{\Omega_c} + \left(\frac{r}{R_v}\right)^3 \right)
\end{flalign}
where $R_v = r_i(1+\delta_i)/(2\delta_i)$ is the virial radius for the dark matter halo. Here, $r$ is co-moving radius of the baryon sphere that by definition
contains all the baryons of mass ($M_b = M_h \Omega_b/\Omega_c$) and the contribution to $M(r)$ due to the uniformly distributed dark matter $(r/R_v)^3 M_h$ decreases.
\end{sloppypar}

\subsection{\underline{Heating due to primordial black holes}}

The rate of accretion of mass onto an individual PBH is taken as the minimum of the Bondi accretion rate and the Eddington accretion rate,
\begin{equation}
{\rm \dot{m}_{PBH}} = \min \left(\frac{\pi G \rho_b}{c_s^3}{\rm m_{PBH}^2} \ , \ \frac{4\pi G m_p}{\sigma_{\rm T} c} {\rm m_{PBH}}\right)
\label{eq:2}
\end{equation}
where $G$ is the gravitational constant, $\rho_b$ is infalling gas density and $c_s$ is the sound speed in the infalling gas, $\sigma_{\rm T}$ is Thompson scattering cross-section, $c$ is speed of light and the $m_p$ is the mass of proton.

The total energy produced per unit comoving volume per unit time by the accretion of mass onto primordial black holes can be given by,
\begin{equation}
\frac{{\rm d}E_{\rm PBH}}{{\rm d}V{\rm d}t} =  \dot{E} \left( \frac{\#PBH}{Volume} \right) = \epsilon \ \frac{\rm \dot{m}_{PBH}} {\rm m_{PBH}} c^2 \Omega_{\text{PBH}} \rho_{\rm{c0}} (1+z)^3
\label{eq:3}
\end{equation}
where $\rho_{c0}$ is the critical density of the Universe today and $z$ is the redshift. 

Accretion theory is replete with various possible scenarios of accretion; the
most apt mode for PBH could be advection dominated disk accretion flow [ADAF, \citet{1995ApJ...452..710N}] or sub-Eddington Bondi accretion \citep{1952MNRAS.112..195B} beside the usual thin disk mode \citep{1973A&A....24..337S}. Based on relativistic theory it
is reasonable to assume an accretion efficiency of $\displaystyle \epsilon = 1 - \sqrt{1-2/(3 r_I)} = 0.06 -0.43 \sim 0.1$ where $r_I$ is the ISCO radius that varies from $6 \rightarrow 1$ as the spin varies from $a=0 \rightarrow 1$. However, the detailed geometry and the accretion mode play a key role in determining $\epsilon$. For example, in case
of thin disks it has been seen from observations and theoretical models that $\epsilon \sim 0.1$ for ${\rm \dot{m}_{PBH}} > {\rm \dot{m_E}}$ \citep{2001ApJ...549..100P}. The sub-Eddington flows are known to be radiatively inefficient \citep{2003MNRAS.343L..59H}. If  the accretion is sub-Eddington, ${\rm \dot{m}_{PBH}} < {\rm \dot{m}_E}$, then the radiative efficiency $\epsilon \propto {\rm \dot{m}_{PBH}}$.   
A conservative estimate considering various allowed modes for PBH accretion has been suggested to take a simple form \citep{2008ApJ...680..829R}; we take it to be $<$ 10\% and assume it to be $\propto \dot{m}$. For the maximum radiative efficiency we have taken a typical value as 10\%. To summarize our assumption,
\begin{equation}
\epsilon = 
\begin{cases} 
0.1, & \frac{\rm \dot{m}_{PBH}}{\rm \dot{m}_E} \geq 1 \\
0.1 \frac{\rm \dot{m}_{PBH}}{\rm \dot{m}_E}, & \frac{\rm \dot{m}_{PBH}}{\rm \dot{m}_E} < 1
\end{cases}
\label{eq:4}
\end{equation}
where ${\rm \dot{m}_E}$ is the Eddington rate.

\subsection{\underline{Thermal evolution of the collapsing gas}}

Our prescription for the thermal evolution of the collapsing gas is based on the prescription given in \citet{2017PhRvD..95h3006C}. We have added an additional term for the accreting primordial black hole heating and we also incorporate cooling (Equation~\ref{eqn:dedvdt}). The evolution of the gas temperature ($T_g$), ionization fraction ($x_e$)and H$_2$ fraction ($x_{\rm H_2}$) is calculated by solving coupled differential Equations~(\ref{eqn:1}{--}\ref{eqn:3}): 

\begin{flalign}
\frac{{\rm d}T_{\rm{g}}}{{\rm d}t} &= -2 \frac{v_r}{r} T_{\text{g}} + \left(\frac{m_p}{m_e} \frac{2\alpha_t}{(1+x_e)}+\frac{x_e^2 \Gamma _{\text{B}}}{1+x_e}\right)\left(T_{\gamma } - T_{\text{g}} \right) - \frac{2 K_h}{3 k_B \left(1 + f_{\text{He}}+x_e\right)} \label{eqn:1} \\ 
\frac{{\rm d} x_e}{{\rm d}t} &= \left[\beta _e(1-x_e) \ e^{\left(\frac{-E_{\alpha }}{k_B T_{\gamma }}\right)}-x_e^2 \alpha _e n_b -\left(I_{\chi \alpha } + I_{\text{$\chi $i}}\right)\right]C  \label{eqn:2}\\
\frac{{\rm d}x_{\rm H_2}}{{\rm d}t} &= k_{form} n_b x_e (1-x_e-2x_{\rm H_2}) - k_{des} n_b x_{\rm H_2}   \label{eqn:3}
\end{flalign}
Here $r$ is the comoving radius of the outer most shell of the gas and $v_r$ is its radial velocity, $T_\gamma$ is the CMB temperature which evolves as $T_\gamma(z) = 2.725 (1+z)$ and $n_b = \rho_b/m_p$ is the baryon number density of the infalling gas. 

The Compton friction $\alpha_t$ is given by, 
\begin{equation}
\alpha _t = \frac{4}{3} \frac{\epsilon_{\gamma} \sigma_T x_e}{m_p c}
\end{equation}
where $\epsilon_{\gamma}$ is the radiation energy density ($a T_\gamma^4$, $a$ is radiation constant). The effective recombination rate and the photoionization rate, $\alpha_e$ and the $\beta_e$ respectively, are given 
by,
\begin{equation}
\alpha _e = 2.84\times10^{-13} \left(\frac{T_{\gamma }}{10^4 \ K} \right)^{-1/2}
\end{equation}
\begin{equation}
\beta _e = \alpha _e \exp \left(-\frac{E_2}{k_B T_{\gamma }}\right) \left(\frac{\sqrt{2 \pi  k_B m_e T_{\gamma }}}{h_p}\right)^3
\end{equation}
where $E_2$ is the energy of the Hydrogen atom at n=2 level, $k_B$ is the Boltzmann constant, $h_p$ is the Planck's constant and $m_e$ is the mass of an electron. The $C$ factor in the Equation~(\ref{eqn:2}) is given by
\begin{equation}
C = \frac{1 +  K \ n_{\text{H}} \ \Lambda _{21} \ (1-x_e)}{1 + K \ \left(\beta _e+\Lambda _{21}\right) \ n_{\text{H}} \ (1-x_e)} \\
\end{equation}
where $n_{\rm H}$ is hydrogen number density, $\Lambda_{21}$ (=8.23) is the decay rate for the transition between the levels 2s to 1s of the hydrogen atom and the parameter $K$ is given by,
\begin{equation}
\quad K = \frac{\lambda _{\alpha }^3}{8 \pi  H}
\end{equation}
where $\lambda_\alpha$ is the wavelength of the Ly-$\alpha$ transition and $E_\alpha$ is the energy corresponding $\lambda_\alpha$, and $H$ is the Hubble parameter. 

\begin{sloppypar}
$\Gamma_B$ is the temperature equilibration time due to Bremsstrahlung absorption and emission given by, 
\begin{equation}
\Gamma _{\text{B}}^{-1} = 4.6 \times 10^9 \left(\frac{T_{\text{g}}}{10^3 \ K}\right)^{1/2} \left(\frac{g_B n_{\text{H}}}{10^3 \ {\rm cm}^{-3}}\right)^{-1}
\end{equation}
where $g_B\approx1.2$ is the averaged Gaunt factor.
\end{sloppypar}

The quantities $I_{\chi\alpha}$, $I_{\text{$\chi $i}}$, and $K_h$ are factors corresponding to the additional energy injection (in our case the PBH heating) which affects the ionization from the ground state, ionization from excited states, and heating of the collapsing gas. The quantity $I_{\chi\alpha}$ is the ionization rate of hydrogen atom from n=2 due to PBH heating, $I_{\text{$\chi $i}}$ is the direct ionization rate of the hydrogen atom, $K_h$ is the fraction of the total energy injection going in to the heating of the medium. Each of these injections is dependent on the injection energy through: 
\begin{flalign}
I_{\chi \alpha } &= \frac{(1-C) \ ({\rm d}E/{\rm d}V{\rm d}t) \ \chi _{\alpha }}{E_{\alpha } \ n_{\text{H}}} \\
I_{\text{$\chi $i}} &= \frac{({\rm d}E/{\rm d}V{\rm d}t) \ \chi _{\rm i}}{E_{\rm i} \ n_{\text{H}}} \\ 
K_h &= \frac{({\rm d}E/{\rm d}V{\rm d}t) \ \chi _h}{n_{\text{H}}}  
\end{flalign}
Here $n_{\rm H}$ is the hydrogen number density, and $E_i$ is the ionization energy for the ground state electron in hydrogen atom and $E_\alpha$ is the difference in binding energy between the n=1 and the n=2 electron energy levels. The quantity $C$ is related to the probability for an excited hydrogen atom to emit a photon prior to being ionized \citep{2017PhRvD..95h3006C}. The quantities $\chi _{\alpha }$, $\chi _{\rm i}$ and $\chi _h$ are efficiencies for energy interactions through each channel, and are taken to be in the following form \citep{2014PhRvD..89j3508M}
\begin{equation}
\chi _{\alpha } = \chi _i ; \quad \chi _{\rm i} = \frac{1}{3} \left(1-x_{\rm H}\right); \quad \chi _h = \frac{1+2 x_{\rm H} + f_{\text{He}} \left(2 x_{\text{He}}+1\right)}{3 \left(f_{\text{He}}+1\right)} 
\label{eq:chi}
\end{equation}
where $x_{\rm H}$ is the ratio of ionized hydrogen to total hydrogen, $x_{\rm He}$ is the ratio of ionized helium to total helium and $f_{He}$ is the helium fraction given by $f_{He} = Y_p/(4(1-Y_p))$ where $Y_p$ is the helium mass fraction (= 0.24). 
In reality, the fraction of energy going in to   ionization of the ambient medium is more complex than the simple linear form given by the Equation~(\ref{eq:chi}), however we use this equation as it 
as a good approximation \citep{2014PhRvD..89j3508M}. 

The total effective energy injection rate per volume is given by 
\begin{equation}
\frac{{\rm d}E}{{\rm d}V{\rm d}t} = \frac{{\rm d}E_{PBH}}{{\rm d}V{\rm d}t} - L_{cool} 
\label{eqn:dedvdt}
\end{equation}
where ${dE_{PBH}}/{dVdt}$ is the heating due to accreting primordial black holes and the $L_{cool}$ is the cooling due to the atomic and molecular hydrogen cooling. 

\subsection{\underline{Atomic (H) and molecular hydrogen (H$_2$) cooling}}

Since the collapsing gas is metal free the only dominant cooling mechanism apart from the atomic hydrogen line cooling and the inverse-Compton process is the molecular hydrogen (H$_2$) cooling. The 
atomic hydrogen cooling is dominant when the temperature of the gas is $\gtrsim 10^4$ K, below this temperature the H${\rm \scriptstyle I}$ line cooling becomes inefficient and H$_2$ cooling takes over. 

To compute the atomic and molecular hydrogen cooling functions ($L_{cool}$) 
we follow the prescription given in \citet{1979ApJS...41..555H} and \citet{1998A&A...335..403G}.

\subsection{\underline{Evolution of molecular hydrogen (H$_2$) fraction}}

In the set up of the high redshift Universe (metal-free gas) the following reactions are the important channels for the formation and destruction of H$_2$ molecules \citep{2010ApJ...721..615S,2008MNRAS.387.1589S}, 

\renewcommand{\arraystretch}{2}

\begin{table*}[h]
\begin{center}
    \begin{tabular}{ | p{3.5cm} | p{10.5cm}  | }     \hline
    {\bf Reaction}                                             & {\bf Reaction rate coefficient $k$ (cm$^3$ s$^{-1}$)}      \\ \hline
    ${\rm H} + e^- \rightarrow {\rm H}^+ + 2e^-$             & $k_1 = 3 \times 10^{-16} \ (T_\gamma/300)^{0.95} \ \exp(-T_\gamma/9320)$           \\ \hline
    ${\rm H} + e^- \rightarrow {\rm H}^- + \gamma$             & $k_9 = 6.775 \times 10^{-15} \ T_{eV}^{0.8779}$           \\ \hline
    ${\rm H}^- + {\rm H} \rightarrow {\rm H_2} + e^-$          & $\!\begin{aligned}[t] k_{10} &= 1.43 \times 10^{-9}  &\text{for} \ T_{eV} \leq 0.1 \\ k_{10} &= \exp\{-20.069 \ + 0.229({\rm ln}T_{eV}) \ + 0.036({\rm ln}T_{eV})^2\} &\text{for} \ T_{eV} > 0.1 \end{aligned}$        \\ \hline
    ${\rm H}^- + {\rm H}^+ \rightarrow 2{\rm H}$               & $k_{13} = 6.5 \times 10^{-9} T_{eV}^{-0.5}$        \\ \hline
    ${\rm H_2 + H} \rightarrow \rm 3 H$                        & $k_{15} = 5.24\times10^{-7} \ T^{-0.485} \ e^{-52000/T}$        \\ \hline
    ${\rm H_2} + {\rm H}^+ \rightarrow {\rm H_2}^+ + {\rm H}$  & $\!\begin{aligned}[t] k_{17} = &\exp\{-24.249 \ + 3.401({\rm ln}T_{eV}) \ - 3.898({\rm ln}T_{eV})^2\} \end{aligned}$        \\ \hline
    ${\rm H_2} + e^- \rightarrow 2{\rm H} + e^-$               & $k_{18} = 5.6 \times 10^{-11} e^{-102124/T} T^{0.5}$        \\ \hline
    ${\rm H}^- + e^- \rightarrow {\rm H} + 2e^-$               & $\!\begin{aligned}[t] k_{19} &= \exp\{-18.018 \ + 2.361({\rm ln}T_{eV}) \ - 0.283({\rm ln}T_{eV})^2\} \end{aligned}$        \\ \hline
    ${\rm H}^- + {\rm H} \rightarrow 2{\rm H} + e^-$           & $\!\begin{aligned}[t] k_{20} &= 2.56 \times 10^{-9} \times T_{eV}^{1.78186} &\text{for} \ T_{eV} \leq 0.1 \\ k_{20} &= \exp\{-20.373 \ + 1.139({\rm ln}T_{eV}) \ - 0.283({\rm ln}T_{eV})^2\} &\text{for} \ T_{eV} > 0.1 \end{aligned}$        \\ \hline
    ${\rm H}^- + {\rm H}^+ \rightarrow {\rm H_2^+} + e^-$      & $\!\begin{aligned}[t] k_{21} &= 4.0 \times 10^{-4} \times T^{-1.4} \times e^{-15100/T} &\text{for} \ T \leq 10^4 \\ k_{21} &= 10^{-8} \ T^{-0.4} &\text{for} \ T > 10^4 \end{aligned}$        \\ \hline
    ${\rm H}^- + \gamma \rightarrow {\rm H} + e^-$             & $k_{\gamma} = 4\left( \frac{2\pi m_e k_B T_\gamma}{h^2} \right)^{3/2} \ \exp\left( \frac{-0.754 eV}{k_B T\gamma} \right) k_1 $    \\ \hline
    \end{tabular}
\end{center}
\caption{Rection rates coefficients \citep{2010MNRAS.402.1249S,2014A&A...561A..13B,2008MNRAS.387.1589S}. In the table $T \equiv T_g$ and $T_{eV}$ is $T_g$ in units of electron-Volts.}
\label{table:1}
\end{table*}

\begin{flalign}
&{\rm H} + e^- \rightleftharpoons {\rm H}^- + \gamma ; \quad {\rm H}^- + {\rm H} \rightleftharpoons {\rm H_2} + e^- \\
&{\rm H} + {\rm H}^+ \rightleftharpoons {\rm H_2}^+ + \gamma ; \quad {\rm H_2}^+ + {\rm H} \rightleftharpoons {\rm H_2} + {\rm H}^+ \\
&\rm 3 H \rightleftharpoons H_2 + H ; \quad
\end{flalign}
Also, the destruction of H$_2$ can be due to the destruction of intermediary ${\rm H}^-$ through following reactions, 
\begin{flalign}
&{\rm H}^- + \gamma \rightarrow {\rm H} + e^- ; \quad {\rm H}^- + {\rm H}^+ \rightarrow 2{\rm H} \\
&{\rm H}^- + e^- \rightarrow {\rm H} + 2e^- ; \quad {\rm H}^- + {\rm H}^+ \rightarrow {\rm H_2^+} + e^-
\end{flalign}

For the H$_2$  formation and destruction rates, we follow \citet{2010ApJ...721..615S} and the references therein. The net rate of formation, $k_{form}$, and destruction, ($k_{des}$) of H$_2$ is taken as 

\begin{flalign}
k_{ form} &= \frac{k_9 k_{10} x_{\rm H{\scriptscriptstyle I}} n_b}{k_{10}  x_{\rm H{\scriptscriptstyle I}} n_b+ k_{\gamma}+(k_{13} + k_{21})x_p n_b + k_{19} x_e n_b + k_{20}x_{\rm H{\scriptscriptstyle I}} n_b} \\
k_{des} &= k_{15} x_{\rm H{\scriptscriptstyle I}} + k_{17}x_p + k_{18} x_e 
\end{flalign}

For all the above reaction rates, we follow the \citet{2010MNRAS.402.1249S} and \citet{2008MNRAS.387.1589S} except for $k_{15}$ which is a three-body reaction (${\rm H_2 + H \rightleftharpoons 3 H }$) rate, for which we follow recent results from \citet{2014A&A...561A..13B}. A list of all these reaction rates is given in the Table~\ref{table:1}, which only shows the leading terms; for the actual fit, please see the references. 

\section{Effect of PBH heating}

\begin{figure}[!ht]
\begin{center}
\includegraphics[width=1.0\columnwidth]{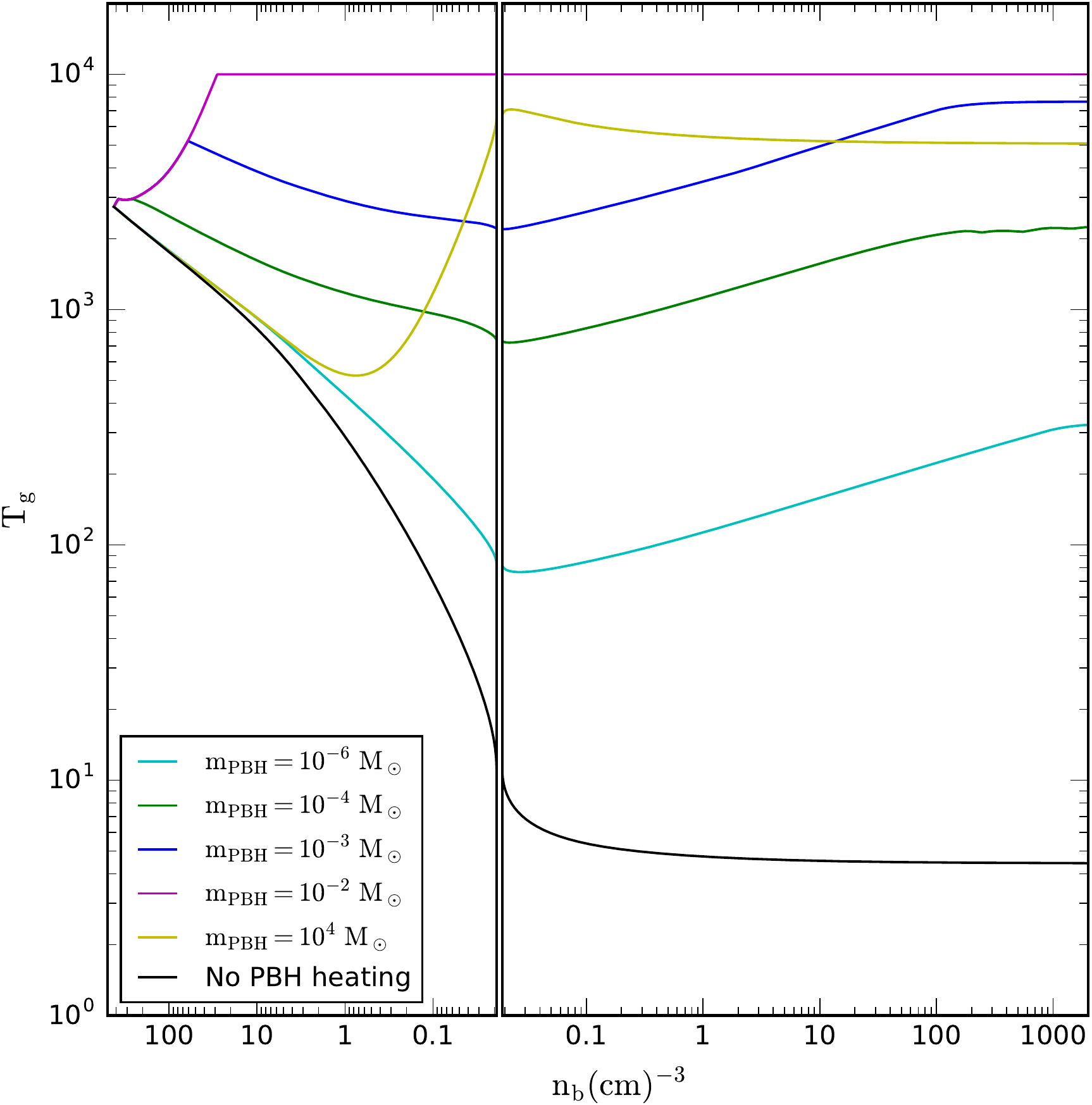}
\caption {The evolution of gas temperature T$_{\rm g}$ with gas density $n_{\rm b}$ as it collapses is shown for various values of the PBH mass m$_{\rm PBH}$. The left panel shows the expansion phase whereas the right panel shows the collapse phase.}
\label{fig:one}
\end{center}
\end{figure}

\begin{figure}[!ht]
\begin{center}
\includegraphics[width=1.0\columnwidth]{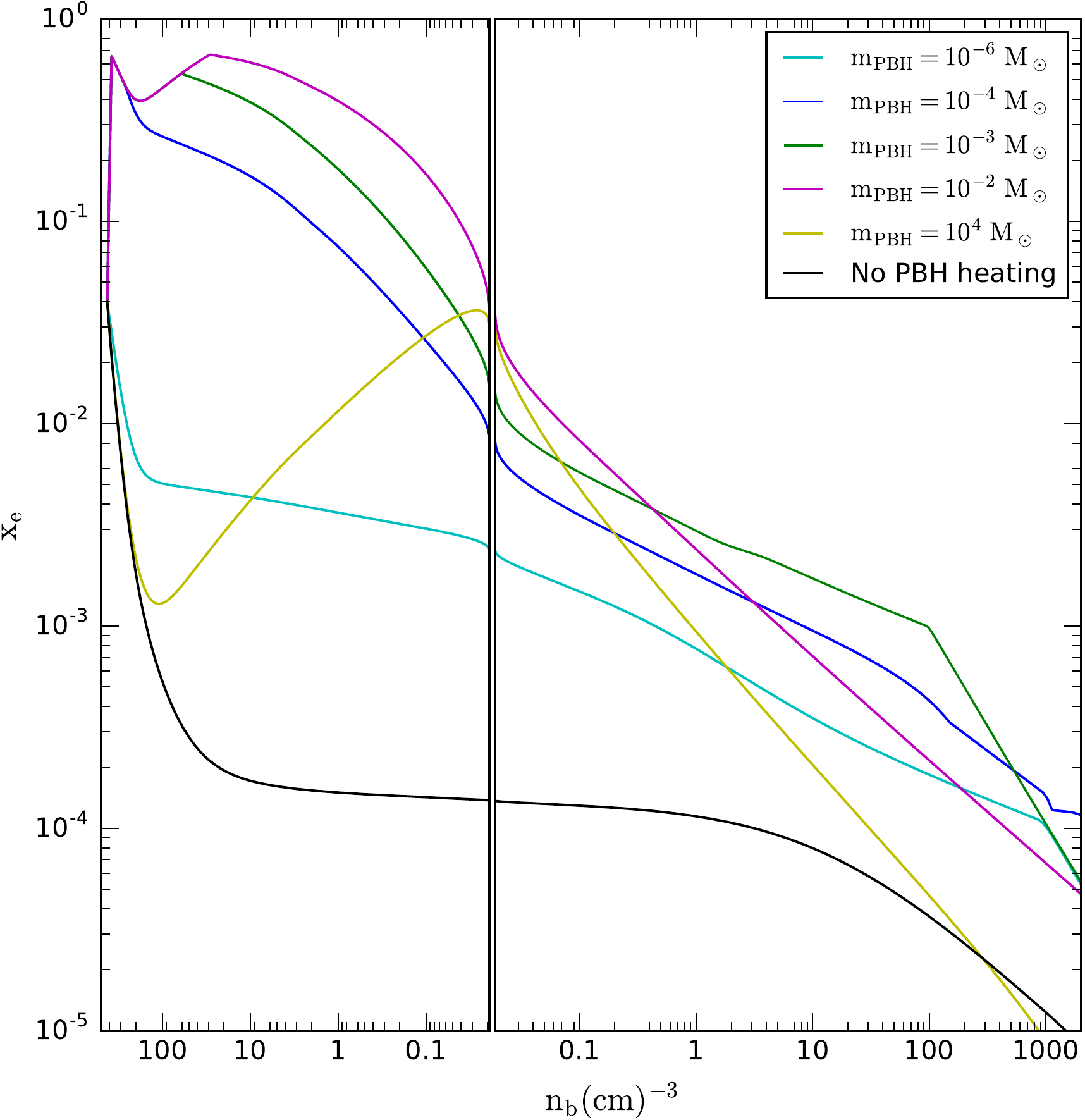}
\caption {The evolution of ionization fraction $x_e$ with gas density $n_{\rm b}$ as it collapses is shown for various values of the PBH mass m$_{\rm PBH}$. The left panel shows the expansion phase whereas the right panel shows the collapse phase.}
\label{fig:two}
\end{center}
\end{figure}

\begin{figure}[!ht]
\begin{center}
\includegraphics[width=1.0\columnwidth]{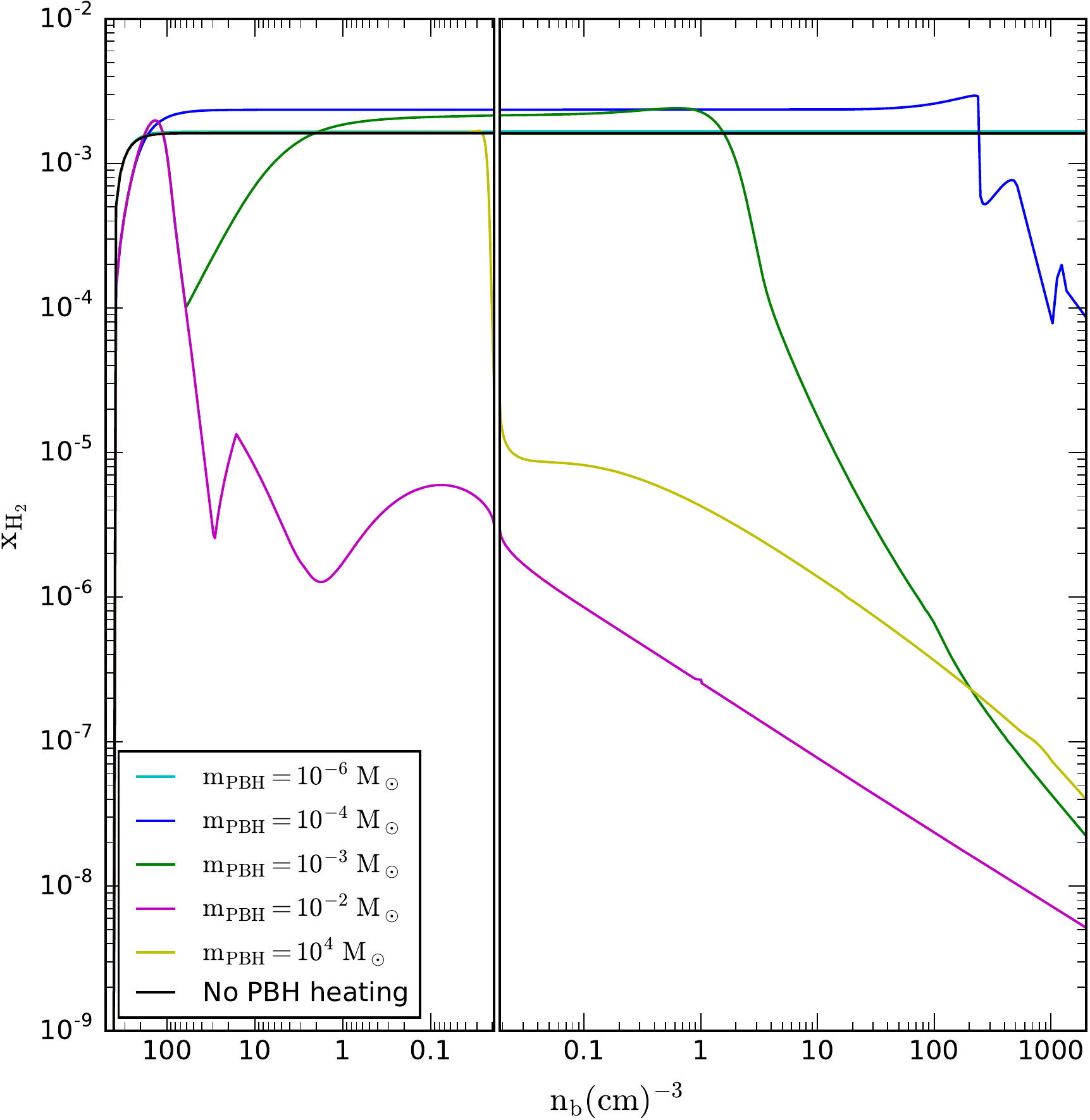}
\caption {The evolution of ${\rm H}_2$ fraction $x_{\rm H_2}$ with gas density $n_{\rm b}$ as it collapses is shown for various values of the PBH mass, m$_{\rm PBH}$. The left panel shows the expansion phase whereas the right panel shows the collapse phase.}
\label{fig:three}
\end{center}
\end{figure}

\begin{figure}[!ht]
\begin{center}
\includegraphics[width=1.0\columnwidth]{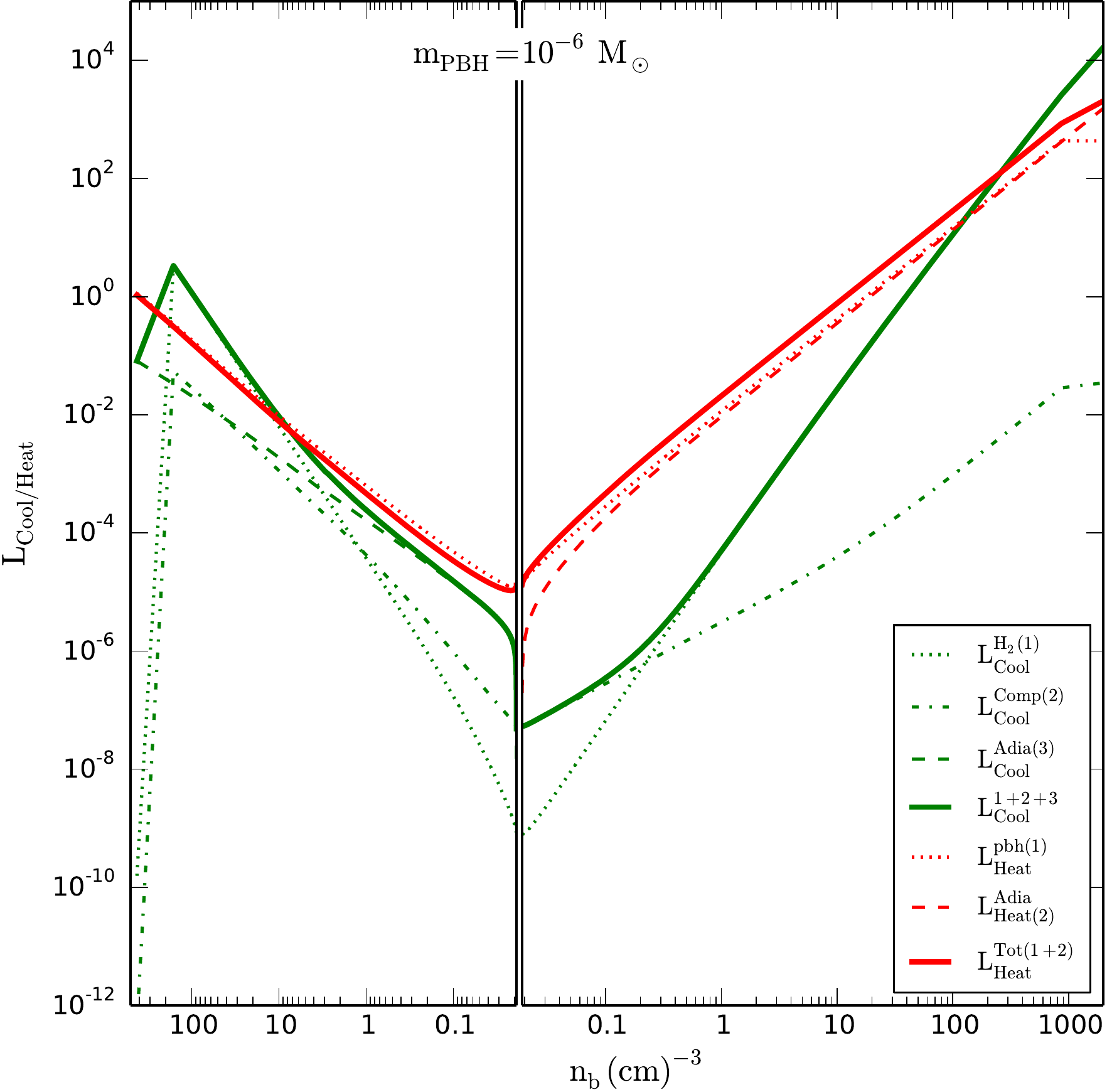}
\caption {Relative strengths of various heating and cooling processes involved (H$_2$-cooling, Compton cooling, adiabatic cooling/heating, PBH heating), for the case of ${\rm m_{PBH} = 10^{-6} M_\odot}$. The rates are in the units of $\sim H_0 dt/dz = \Omega_m^{-1/2}(1+z)^{-5/2}$.}
\label{fig:four}
\end{center}
\end{figure}

\begin{figure}[!ht]
\begin{center}
\includegraphics[width=1.0\columnwidth]{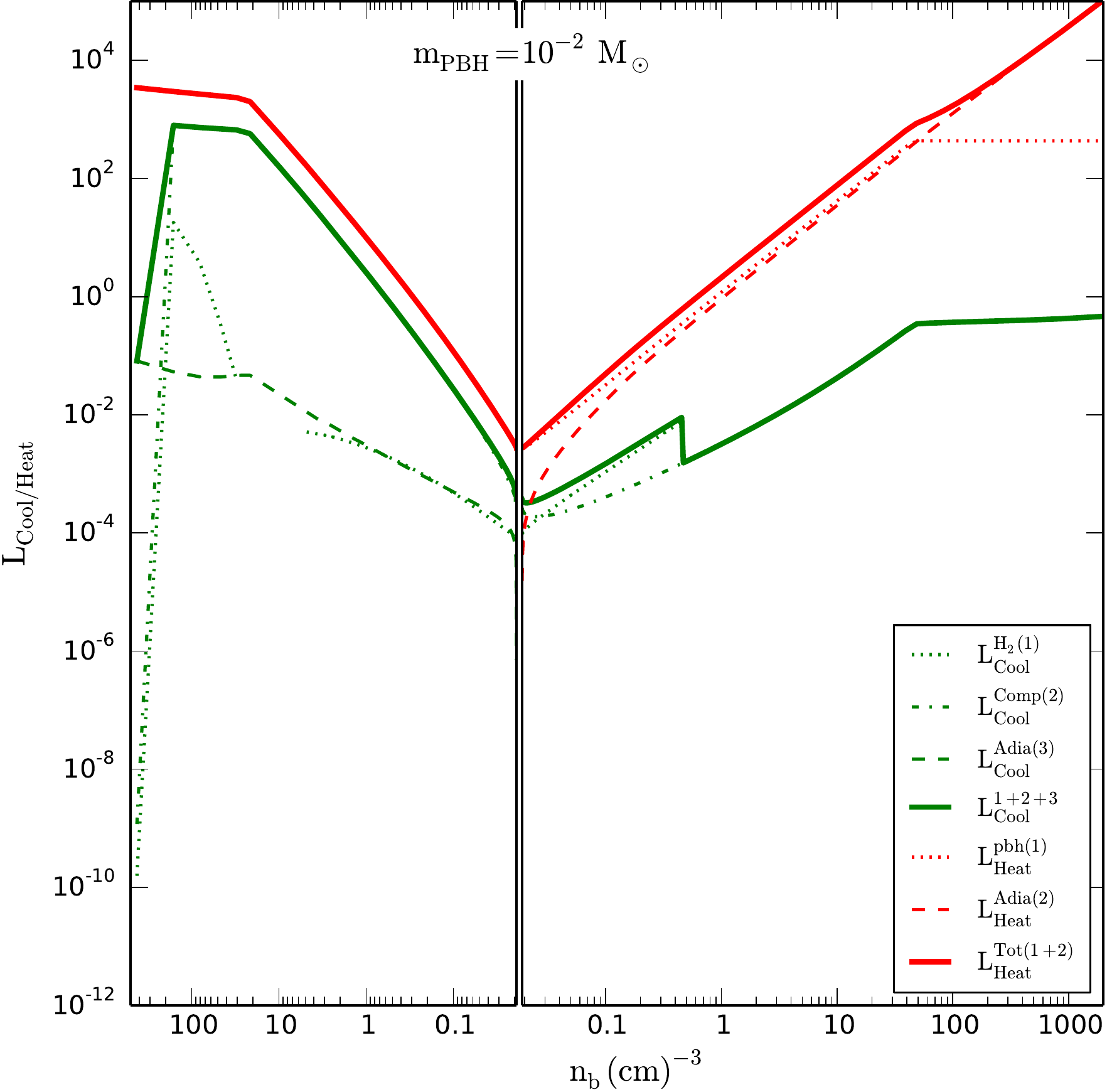}
\caption {Relative strengths of various heating and cooling processes involved (H$_2$-cooling, Compton cooling, adiabatic cooling/heating, PBH heating), for the case of ${\rm m_{PBH} = 10^{-2} M_\odot}$. The rates are in the units of $\sim H_0 dt/dz = \Omega_m^{-1/2}(1+z)^{-5/2}$.}
\label{fig:five}
\end{center}
\end{figure}

The Figures~\ref{fig:one}-\ref{fig:three} show the evolution of the temperature, $T_g$, the ionization fraction, $x_e$,  and the H$_2$ fraction, $x_{\rm H_2}$. The time evolution in these figures is towards the right. The gas density falls (as the halo expands with the background Universe) until the turnaround, and then increases again as the halo collapses. The figures show the interplay between several physical effects. The accretion rate for a higher mass PBH is more and therefore the heating also increases with the mass of the PBHs. The PBH heating increases the temperature of the gas which increases the ionization fraction due to enhanced collisional ionization. However, the increase in temperature play a complicated role in the evolution $x_{\rm H_2}$; it increases the collisional destruction rate of H$_2$ but at the same time an increase in ionization fraction, $x_e$, tends to increase the formation rate of H$_2$ competing with its collisional dissociation. H$_2$ cooling depends on the temperature as well as the molecular hydrogen fraction. 

Figures~\ref{fig:four}-\ref{fig:five} show the relative strength of various heating and cooling mechanisms involved for the case of ${\rm m_{PBH} = 10^{-6} M_\odot}$ and ${\rm m_{PBH} = 10^{-2} M_\odot}$ respectively. We can clearly see from these figures that for lower mass (${\rm 10^{-6} M_\odot}$) PBH heating, H$_2$-cooling eventually wins, whereas in the case of sufficiently higher mass (${\rm 10^{-2} M_\odot}$) PBH heating is high enough to be able to suppress the H$_2$ cooling. In this case, the temperature of the collapsing gas can rise until it reaches $10^4$ K, beyond which the H{\scriptsize I} line cooling takes over, which is an efficient cooling process for the collapsing gas which is optically thin, and therefore the temperature of the collapsing gas does not rise above the $10^4$ K.

With the low mass PBH heating, the H$_2$ formation increases very rapidly and results in a lowering of temperature due to H$_2$ cooling which in turn drop the ionization fraction. But for high enough heating rate H$_2$ dissociation finally takes over and the H$_2$ fraction drops rapidly in the collapsing phase of the gas; thus the H$_2$ cooling become weak. At the same time, as the halo gas collapses further, in the high density environment aided with the high temperature, the three body destruction rate of H$_2$ becomes important and the H$_2$ fraction goes down even more rapidly. Once a critical density reaches a value $n_b \gtrsim 1000$ per cc, the roto-vibrational states of H$_2$ molecules reaches a local thermodynamic equilibrium and H$_2$ cooling becomes inefficient. Therefore the further gas collapse is not affected by the H$_2$ cooling and the temperature of the gas remains near $\sim 10^4$ K. 

We carried out the calculation for higher values of PBH masses and found that for any value of PBH mass above $ \sim 10^{-2} \ {\rm M}_\odot$ it is possible to heat the collapsing gas enough to avoid the H$_2$ cooling. Though for very high mass ($\gtrsim 10^{4} \ {\rm M}_\odot$) PBHs the allowed abundance limit is too low to produce enough heating (see Figure~\ref{fig:one}). 

\newpage

\section{Further collapse}

In absence of $\rm H_2$ cooling the metal-free primordial gas collapses quasi-isothermally. 
The high Jeans mass ($\sim 10^5 \ {\rm M}_\odot$) due to the high temperature ($\sim 10^4$ K) inhibits fragmentation and also ensures gas accretion onto the central proto-star at a very high rate, $\dot{M} \sim 1 {\rm M}_\odot \ {\rm yr}^{-1}$. The result of this could be either of two scenarios, {\it (i)} the collapsing gas with a little initial angular momentum turns into a supermassive star (SMS) $M \sim 10^5 \ {\rm M}_\odot$, or {\it (ii)} if the initial angular momentum is high, a rotationally supported massive accretion disk forms around a relatively less massive ($\sim 100 \ {\rm M}_\odot$) proto-star at the centre. Both the cases result in a massive black hole ($M_\bullet \sim 10^5 \ {\rm M}_\odot$) at the centre within a Myr time scale, which subsequently grows into a high redshift SMBH. 
Since simulations indicate the Peeble's parameter $\lambda \sim 0.07$, it is
likely that the final collapse occurs through a supermassive disk phase in most
cases without a supermassive star (see Figure~\ref{flowchart}).

\tikzstyle{block} = [rectangle, draw, fill=brown!20,
    text width=18em, text centered, rounded corners, minimum height=2.5em]
\tikzstyle{subblock} = [rectangle, draw, fill=brown!20, 
    text width=9.6em, text centered, rounded corners, minimum height=5em]
\tikzstyle{block_nb1} = [rectangle, draw=none, fill=none, 
    text width=15em, text centered, rounded corners, minimum height=1.0em]
\tikzstyle{block_nb2} = [rectangle, draw=none, fill=cyan!20, 
    text width=9.0em, text centered, rounded corners, minimum height=1.0em]

\begin{figure}[h]
\centering
\hspace{-1.1cm}
\begin{tikzpicture}[]
    \node [block] (0) {$M_h \sim 10^8 \ {\rm M}_\odot$, $T_v \sim 10^4$ K, \\ $z_{v} \sim 15-20, \ R_v \sim 1 kpc$};
    \node [block_nb1, below=1em of 0] (1) {\scriptsize Rapid collapse without fragmentation, \\ and with high in fall speed $\sim c_s$ \\ \vspace{-0.4em} owing to high $T_v$ and deep potential well};
    \node [block_nb2, below left=2em and -7em of 1] (2a) {\scriptsize little \\ \vspace{-0.5em} angular moentum};
    \node [block_nb2, below right=2em and -7em of 1] (2b) {\scriptsize with \\ \vspace{-0.5em} angular moentum};
    \node [block_nb1, below=1em of 2a] (2c) {\scriptsize within a few Myrs of \\ \vspace{-0.5em} rapid accretion phase};
    \node [block_nb1, below=1em of 2b] (2d) {\scriptsize within a few Myrs of \\ \vspace{-0.5em} rapid accretion phase};
    \node [subblock, below=1em of 2c] (3a) {Supermassive star; $M_s \sim 10^{4-6} \ {\rm M}_\odot$ \\ $r_s \lesssim  0.1 pc$};
    \node [subblock, below=1em of 2d] (3b) {Supermassive disk; $M_d \sim 10^{4-6} \ {\rm M}_\odot$ \\ $r_d \lesssim  10 pc$};
    \node [block_nb1, below right=2em and -7.2em of 3a] (4) {\scriptsize Accretion within a few Myr \\ \vspace{-0.5em} due to gravitational instabilities};
    \node [block, below=1em of 4] (5) {Intermediate mass black hole; $M_\bullet \sim 10^{4-6} \ {\rm M}_\odot$};
    \node [block_nb1, below=1em of 5] (6) {\scriptsize within 400 - 500 Myr \\ \vspace{-0.5em} through Eddington limited accretion};
    \node [block, below=1em of 6] (7) {Supermassive black hole (SMBH); $M_\bullet \sim 10^{8-9} \ {\rm M}_\odot, \ z \sim 6-7$};

    \draw[->,thick,line width=0.5mm,draw=gray!80] (0) to (1);
    \draw[->,thick,line width=0.5mm,draw=gray!80] (1.south) to [out=270,in=90] (2a.north);
    \draw[->,thick,line width=0.5mm,draw=gray!80] (1.south) to [out=270,in=90] (2b.north);
    \draw[->,thick,line width=0.5mm,draw=gray!80] (2a) to (2c);
    \draw[->,thick,line width=0.5mm,draw=gray!80] (2b) to (2d);
    \draw[->,thick,line width=0.5mm,draw=gray!80] (2c) to (3a);
    \draw[->,thick,line width=0.5mm,draw=gray!80] (2d) to (3b);
    \draw[->,thick,line width=0.5mm,draw=gray!80] (3a.south) to [out=270,in=90] (4.north);
    \draw[->,thick,line width=0.5mm,draw=gray!80] (3b.south) to [out=270,in=90] (4.north);
    \draw[->,thick,line width=0.5mm,draw=gray!80] (4) to (5);
    \draw[->,thick,line width=0.5mm,draw=gray!80] (5) to (6);
    \draw[->,thick,line width=0.5mm,draw=gray!80] (6) to (7);
\end{tikzpicture}
\caption {Flowchart of the proposed DCBH model for the formation of SMBH at high redshift}
\label{flowchart}
\end{figure}
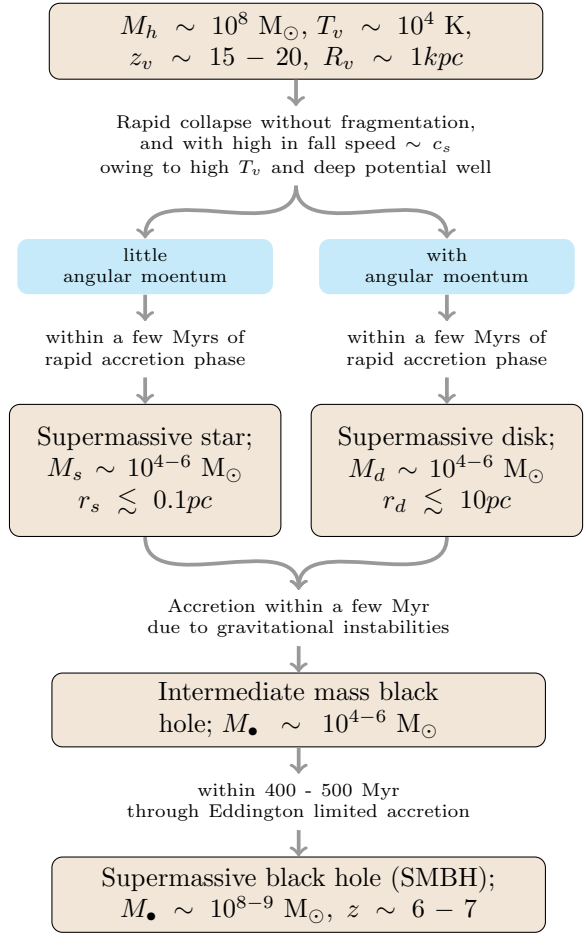

\subsection{\sf Angular momentum transfer}
\begin{sloppypar}
The case when a rotationally supported disk forms around a proto-star, the angular momentum transfer has to be very efficient for it to turn into a massive black hole within the stipulated time scale. Previous studies show that the angular momentum transfer due to $\alpha$-viscosity is very ineffective. The major sources of angular momentum transfer could be the radiative viscosity, radiative drag and gravitational instabilities \citep[and the references therein]{2001A&A...379.1138M}. We briefly discuss these two scenarios in the following sections.
\end{sloppypar}

\subsubsection{\underline {CBR drag}}
\begin{figure}
\begin{center}
\includegraphics[width=1.0\columnwidth]{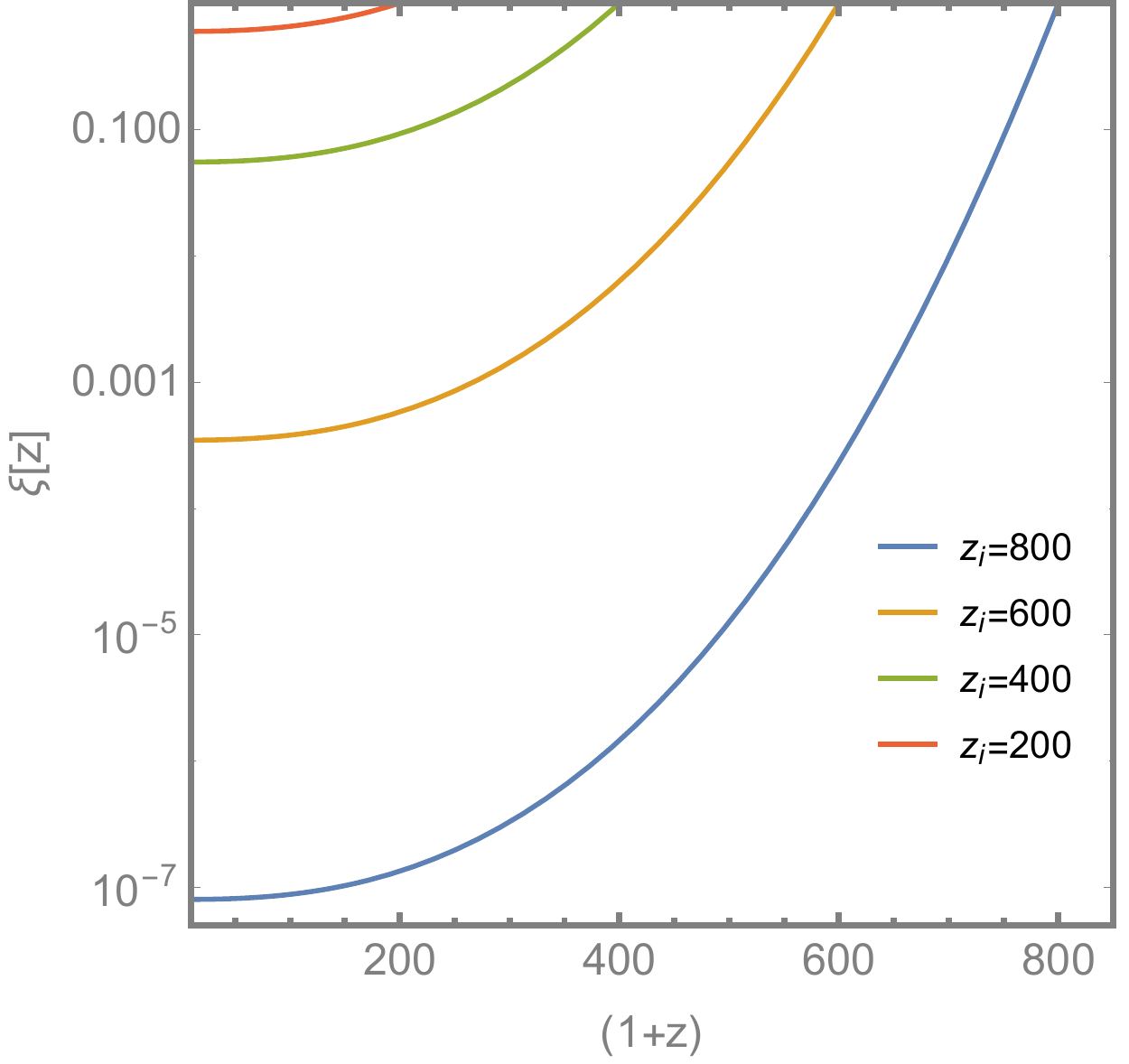}
\caption {A plot of the collapse factor, $\xi(z) = r(z)/r(z_i)$, due to CBR drag at high redshifts.}
\label{fig:col1}
\end{center}
\end{figure}

\begin{figure}
\begin{center}
\includegraphics[width=0.94\columnwidth]{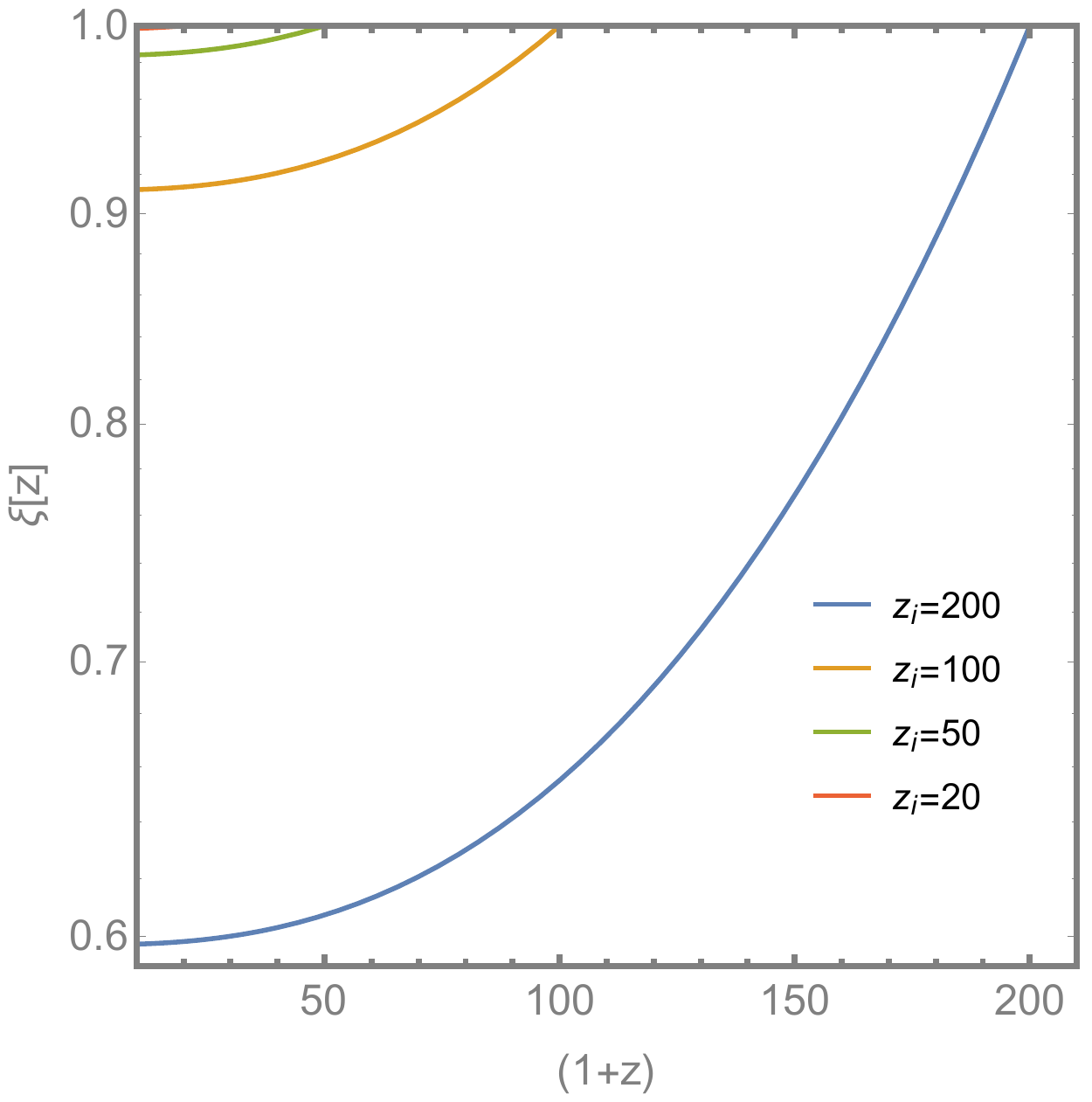}
\caption {A plot of the collapse factor, $\xi(z) = r(z)/r(z_i)$, due to CBR drag at low redshifts.}
\label{fig:col2}
\end{center}
\end{figure}

Cosmic background radiation (CBR) drag due to Compton friction can play an effective role in removing the angular momentum from a highly ionized ($x\approx1$) disk at high redshifts (see Figure~\ref{fig:col1}). A self-similar collapse of a highly ionized ($x\approx1$) Mestel disk due to the Compton drag of CBR is given by [\citet{2003BASI...31..207M}; suitably scaled], 
\begin{small}
\begin{equation}
\xi(z) = \exp\left[- \ 0.4\times10^2 \left(\frac{1+z}{1000}\right)^{5/2} \left(1-\left(\frac{1+z}{1+z_i}\right)^{5/2}\right) \right]
\end{equation}
\end{small}
where $z_i$ is the redshift of the formation of the (ionized) disk. We see that the CBR drag is very effective at high redshifts and can give rise to huge collapse at high redshifts, but if the disk is formed at low redshifts ($z \lesssim 200$) then the CBR drag may not play any important role in collapsing the disk due to inefficient angular momentum transfer (Figure~\ref{fig:col2}). 

For the CBR drag to work as an efficient mechanism for angular momentum transfer, the collapse redshift of the dark matter halo should be very high $\gtrsim 400$. At these redshifts, in a scenario with  $M_h \sim 10^{4-5} \ {\rm M}_\odot$, it may be statistically possible in low spin systems with efficient cooling without fragmentation to produce a supermassive disk which can collapse by CBR drag \citep{2003BASI...31..207M}. 

\subsubsection{\underline{Gravitational instabilities}}
Once a supermassive disk forms, gravitational instabilities can be effective means of angular momentum transport for the further collapse in the DCBH scenario occurring at $z=9-15$. 
A rotating disk can become unstable under its own gravity as soon as the Toomre parameter goes less than 1 \citep{1964ApJ...139.1217T}, 

\begin{equation}
Q = \frac{c_s \omega}{\pi G \Sigma} < 1
\end{equation}
where $c_s$ is the sound speed, $\omega$ is the rotation speed and $\Sigma$ is the surface density of the disk, and $G$ is the gravitational constant. The corresponding time scale of formation of a central massive black hole can be written as \citep{2001A&A...379.1138M}, 

\begin{small}
\begin{eqnarray}
t_g &=& \frac {\omega M} {2 \pi \Pi^g_{r\phi}} \nonumber \\ 
    &=& 4 \ {\rm Myr} \ \cdot \ {\cal M}_{10}^{1/2} \left( \frac {15} {1+z_c}  \right) \left( \frac {0.1} {f_b}  \right) \left( \frac {0.1} {\alpha_g} \right) m_{\bullet 8}
\end{eqnarray}
\end{small}
where ${\cal M}_{10}$ is the mass of the halo in units of $10^{10} \ {\rm M}_\odot$, $\Pi^g_{r\phi}$ is the viscous stress due to gravitational instabilities, $z_c$ is the redshift of collapse (formation of the disk), $m_{\bullet 8}$ is the black hole mass in  units of $10^8 \ {\rm M}_\odot$ and $f_b = \Omega_b/\Omega_m$.

Such collapse times scales are applicable for black hole formation via supermassive disks even with some initial angular momentum, by $z \sim 7$, which may be obtained through the process of DCBH described above. Further detailed investigation of the physics and demographics of such black holes will be presented  in a future work.

\section{Conclusions}

We use the most recent bound on the accreting primordial black holes, and adopt a simple model for the  heating produced by these accreting primordial black holes, and we find that the accreting primordial black holes of masses $\gtrsim 10^{-2} \ {\rm M}_\odot$ could be a potential source of heating in a direct collapse black hole scenario.

We take a simple approach to quantify the effect of possible primordial black holes comprising a fraction of dark matter, on the DCBH scenario of the formation of super massive black holes. We take reasonable  mass-accretion rates and the radiative efficiency of the accretion, but at the same time, we have taken caution to make our calculation as conservative as possible so that we do not over estimate the effect. Also, we have used a simple model for the dynamical collapse of the gas. Nonetheless, our calculation shows that possible accreting primordial black holes of masses around $\lesssim 10 \ {\rm M}_\odot$ could be another potential source of heating in the context of the thermodynamical evolution of the galaxies and their central black holes.

Though our model requires a good fraction of dark matter to be made up of primordial black holes, we argue here that this possibility is not yet ruled out by the current observational constraints on primordial black hole abundances. Moreover, the fact that no dark matter particles have been observed by the world’s most sensitive direct-detection experiments till now \citep{2017PhRvL.118b1303A,2017PhRvL.119b181302A,2017PhRvL.119b181301A}, casts a doubt on the particle physics based dark matter models. 
Though, we have strong constraints on the PBH abundances coming from various observations, most of these constraints are applied to a certain mass window and are based on the assumption that the PBH mass function is monochromatic which is unrealistic. With an extended mass function for the PBH it may be possible to 
explain all the dark matter, even if the density in any particular mass band is small and within the observational bounds [a more detailed discussion on this topic can be found in \citet{2016PhRvD..94h3504C, 2017PhRvD..96b3514C, 2017PhRvD..95h3508K}]. Nonetheless, the monochromatic bounds provide a platform to carry out our calculations to make a conservative estimate about the possible effects of primordial black holes during the phase of early structure formation.

Though the current constraints on PBH are not very robust and should be taken as an order of magnitude estimate, the upcoming data from experiments such as the release of final Planck high $\ell$ data and results from advanced LIGO experiments can play an important role in boosting or ruling out the PBH hypothesis of the dark matter. Also, future and current searches [such as \citet{2017PhRvL.118x1101G}] of possible PBHs around galactic ridge region in the Milky Way using careful measurements of radio and X-ray emission coming from the accretion of gas onto the population of PBHs can be a key to establishing or debunking the PBH hypothesis. 

In the future, we plan to carry out a detailed analysis, using a well-motivated extended mass function for the primordial black hole mass distribution, and also incorporating an efficient mechanism of angular momentum transfer in the collapse physics.

\section*{Acknowledgement}
We thank the referees for useful comments. KLP thanks Shiv Sethi for many useful discussions.

\newpage


\end{document}